\documentclass[twoside]{article} 
\usepackage[american]{babel}
\usepackage{calc}
\usepackage{amsmath}
\usepackage{epsfig}
\usepackage{psfig}
\usepackage{cite}
\begin{document}
\thispagestyle{empty}
\begin{titlepage}
\thispagestyle{empty}
\title{\begin{flushright}{\large  DESY--00-40}\end{flushright}
\vspace*{2cm}
{\bf \LARGE Measurement of Azimuthal Asymmetries \\
in Deep Inelastic Scattering\\}}
\author{ZEUS Collaboration}
\date{ }
\maketitle
\begin{abstract}
\thispagestyle{empty}
The distribution of the azimuthal angle for the charged hadrons has been
studied in the hadronic centre-of-mass system for neutral current
deep inelastic positron-proton scattering with the ZEUS detector at HERA.
Measurements of the dependence of the moments of this
distribution on the transverse momenta of the charged
hadrons are presented.
Asymmetries that can be unambiguously attributed to perturbative
QCD processes have been observed for the first time.
\end{abstract}
\end{titlepage}
\setcounter{page}{1}
\pagenumbering{arabic}
%
%
%
%
\topmargin-1.cm                                                                                    
\evensidemargin-0.3cm                                                                              
\oddsidemargin-0.3cm                                                                               
\textwidth 16.cm                                                                                   
\textheight 680pt                                                                                  
\parindent0.cm                                                                                     
\parskip0.3cm plus0.05cm minus0.05cm                                                               
\def\3{\ss}                                                                                        
\newcommand{\address}{ }                                                                           
\pagenumbering{Roman}                                                                              
                                                   %
\begin{center}                                                                                     
{                      \Large  The ZEUS Collaboration              }                               
\end{center}                                                                                       
  J.~Breitweg,                                                                                     
  S.~Chekanov,                                                                                     
  M.~Derrick,                                                                                      
  D.~Krakauer,                                                                                     
  S.~Magill,                                                                                       
  B.~Musgrave,                                                                                     
  A.~Pellegrino,                                                                                   
  J.~Repond,                                                                                       
  R.~Stanek,                                                                                       
  R.~Yoshida\\                                                                                     
 {\it Argonne National Laboratory, Argonne, IL, USA}~$^{p}$
% ...
\newpage                                                                                           
% ...
                                                           %
                                                           %
\newpage   
                                                           %
                                                           %
\begin{tabular}[h]{rp{14cm}}                                                                       
$^{a}$ &  supported by the Natural Sciences and Engineering Research                               
          Council of Canada (NSERC)  \\                                                            
$^{b}$ &  supported by the FCAR of Qu\'ebec, Canada  \\                                            
$^{c}$ &  supported by the German Federal Ministry for Education and                               
          Science, Research and Technology (BMBF), under contract                                  
          numbers 057BN19P, 057FR19P, 057HH19P, 057HH29P, 057SI75I \\                              
$^{d}$ &  supported by the MINERVA Gesellschaft f\"ur Forschung GmbH, the                          
German Israeli Foundation, the Israel Science Foundation, the Israel                               
Ministry of Science and the Benozyio Center for High Energy Physics\\                              
$^{e}$ &  supported by the German-Israeli Foundation, the Israel Science                           
          Foundation, the U.S.-Israel Binational Science Foundation, and by                        
          the Israel Ministry of Science \\                                                        
$^{f}$ &  supported by the Italian National Institute for Nuclear Physics                          
          (INFN) \\                                                                                
$^{g}$ &  supported by the Japanese Ministry of Education, Science and                             
          Culture (the Monbusho) and its grants for Scientific Research \\                         
$^{h}$ &  supported by the Korean Ministry of Education and Korea Science                          
          and Engineering Foundation  \\                                                           
$^{i}$ &  supported by the Netherlands Foundation for Research on                                  
          Matter (FOM) \\                                                                          
$^{j}$ &  supported by the Polish State Committee for Scientific Research,                         
          grant No. 112/E-356/SPUB/DESY/P03/DZ 3/99, 620/E-77/SPUB/DESY/P-03/                      
          DZ 1/99, 2P03B03216, 2P03B04616, 2P03B03517, and by the German                           
          Federal Ministry of Education and Science, Research and Technology (BMBF)\\              
$^{k}$ &  supported by the Polish State Committee for Scientific                                   
          Research (grant No. 2P03B08614 and 2P03B06116) \\                                        
$^{l}$ &  partially supported by the German Federal Ministry for                                   
          Education and Science, Research and Technology (BMBF)  \\                                
$^{m}$ &  supported by the Fund for Fundamental Research of Russian Ministry                       
          for Science and Edu\-cation and by the German Federal Ministry for                       
          Education and Science, Research and Technology (BMBF) \\                                 
$^{n}$ &  supported by the Spanish Ministry of Education                                           
          and Science through funds provided by CICYT \\                                           
$^{o}$ &  supported by the Particle Physics and                                                    
          Astronomy Research Council \\                                                            
$^{p}$ &  supported by the US Department of Energy \\                                              
$^{q}$ &  supported by the US National Science Foundation                                          
\end{tabular}                                                                                      
                                                           %
                                                           %
\newcommand{\sgeq}
{\mbox{\raisebox{-0.4ex}{$\;\stackrel{>}{\scriptstyle 
\sim}\;$}}}
\newcommand{\sleq}
{\mbox{\raisebox{-0.4ex}{$\;\stackrel{<}{\scriptstyle 
\sim}\;$}}}
\newcommand{\tttld}
{\mbox{\raisebox{0.4ex}{\boldmath$\scriptscriptstyle\sim$}}}
\newpage
\setcounter{page}{1}
\pagenumbering{arabic}
\section{Introduction}
\label{sect:intro}

\begin{figure}[ph!]
\centering
\begin{center}
\mbox{\epsfig{file=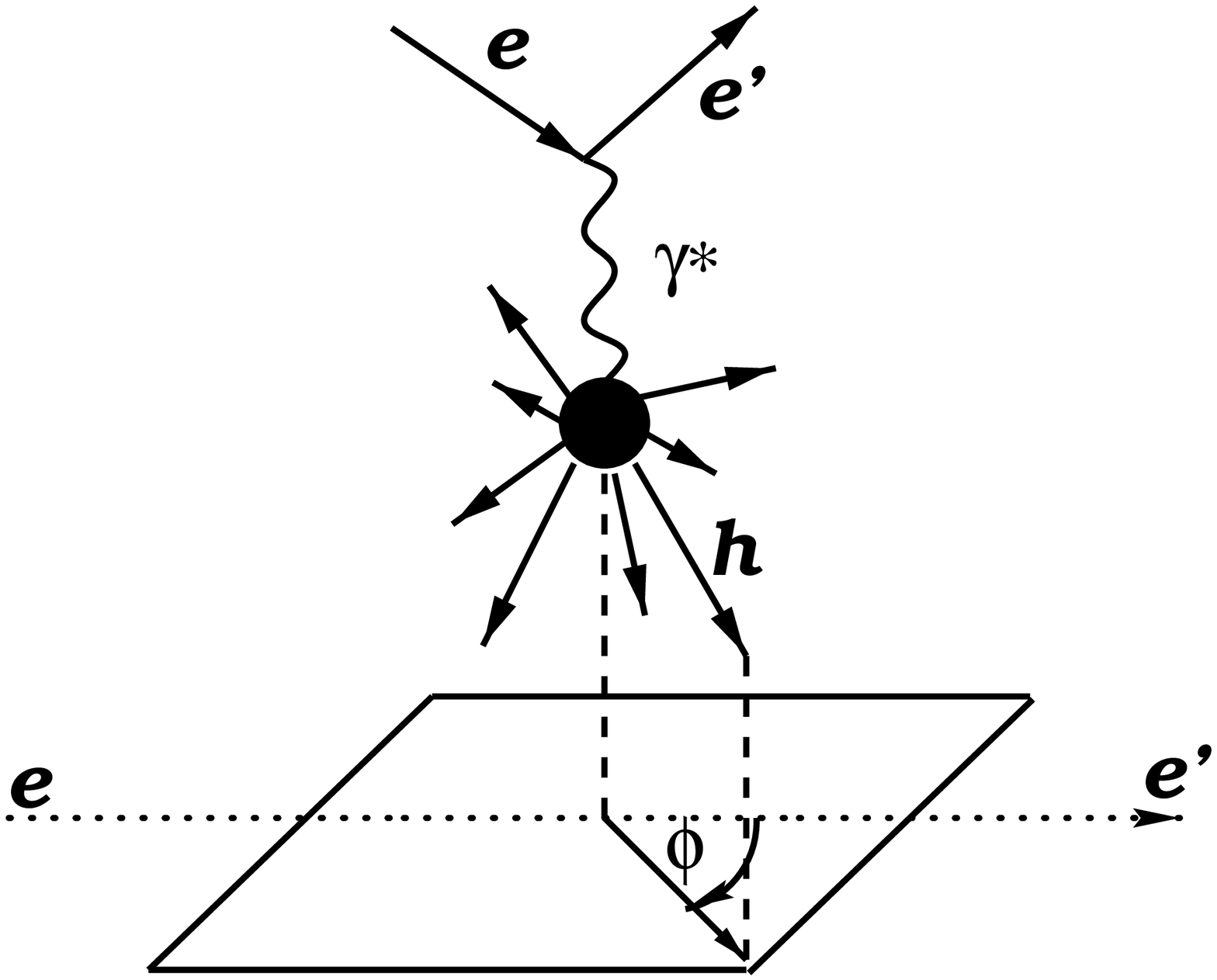,width=\textwidth}}
\end{center}
\caption{
Definition of the azimuthal angle $\phi.$ The incoming lepton
is denoted by $e,$ the scattered  lepton by $e^{\prime},$ the exchanged
boson by $\gamma^*$ and the outgoing hadrons or partons by $h.$ The
dotted line represents the intersection of the $e-e^{\prime}$ scattering plane
with the transverse plane.}
\label{fig:laurel}
\end{figure}

Semi-inclusive processes in deep inelastic scattering (DIS) are
of importance because they can be used to test the
perturbative Quantum Chromodynamic (QCD) description of hadron production
via parton fragmentation.
An observable of particular interest is the distribution
of the azimuthal angle, $\phi,$ (measured in the hadronic centre-of-mass
frame, HCM) between the lepton scattering
plane, defined by the incoming and outgoing lepton momenta,
and the hadron production plane, defined by the exchanged
virtual boson  and an outgoing hadron (Fig.~\ref{fig:laurel}).

Asymmetries in the $\phi$ distribution, i.e. terms proportional to
$\cos \phi$ and $\cos 2\phi$ 
arise whenever a non-zero transverse momentum, in the HCM frame,
is present in the
scattering process.
Consequently both non-perturbative and perturbative QCD
effects~\cite{politzer,cahn,kopp,mendez,chay,ogan,gehrmann} 
give rise to these
asymmetries.  The azimuthal dependence  of parton production has the
form

\begin{equation}
    d\sigma/d\phi  = A + B \cos\phi + C \cos 2\phi .
\label{eq:azi}
\end{equation}

\noindent This form results from the polarisation of the exchanged
virtual boson.
The coefficients $B$ and $C$ depend on the helicities
of the final-state parton(s), partonic transverse momenta
and on colour coherence~\cite{chay}.
The
$\cos 2\phi$ term is expected from interference of amplitudes arising
from the $+1$ and $-1$
helicity components of the transversely-polarised part of the exchanged
boson,
whereas transverse/longitudinal interference gives rise to the
$\cos\phi$ term.

The asymmetry that results from the intrinsic momentum of a quark in the proton
is referred to as the non-perturbative asymmetry.
Since the intrinsic momentum is small, this asymmetry should
fall rapidly with increasing transverse momentum  of
the measured hadron and with
increasing $Q^2$~\cite{cahn},
where
$Q^2\equiv -q^2$
is the negative square of the four-momentum of the virtual exchanged boson.
In particular, the  term $C$ is small at high $Q^2$ for the
non-perturbative effect.

In contrast, the asymmetry associated with leading-order terms in perturbative
QCD calculations, the perturbative asymmetry, is weakly dependent on $Q^2$ and 
persists at high transverse momenta.
The perturbative QCD contribution to
terms $B$ and $C$ is large 
at leading order (LO) in $\alpha_s.$ At this order,
two processes contribute to DIS:
QCD-Compton scattering (QCDC, $\gamma^*q \rightarrow qg$) and boson-gluon
fusion (BGF, $\gamma^*g \rightarrow q \bar q$).
For the QCDC process the scattered quark preferentially appears
 at $\phi$ near
$180^{\circ}$ whilst for the BGF process the $\phi$ dependence has a
symmetry about $180^{\circ},$ as a consequence of the
quark/anti-quark final state.
The BGF process is the dominant contribution to the $\cos 2\phi $ term,
while the QCDC contribution 
 dominates the $\cos \phi $ term.
The $\cos 2\phi $ term can be unambiguously
attributed to perturbative QCD processes in the high $Q^2$ region under
investigation. 

Terms in $\sin\phi$ and $\sin
2\phi$ are also theoretically expected only for deep inelastic scattering with 
polarised leptons or charged current reactions. Even in these cases, their 
magnitude is however estimated to be substantially smaller than terms in 
$\cos\phi$ and $\cos 2\phi$~\cite{gehrmann}.

The form of Eq.~\ref{eq:azi} is expected to be maintained for single
particle production~\cite{chay,gehrmann}, since high-momentum hadrons are
produced close to the direction of the parton.
Measurable perturbative and non-perturbative
predictions for the asymmetries in hadron production require the use of
an appropriate fragmentation function or hadronisation mechanism.
Note that in order to observe the $\cos \phi$ asymmetry, a selection
procedure is required which consistently associates leading
hadrons produced from
{\rm either} quarks {\rm or} gluons~\cite{chay,gehrmann}. 
This is described in Section~\ref{sec:expt}.
The transverse momenta arising from the fragmentation itself does not contribute
to the asymmetry but only smears the observed distribution~\cite{chay}.

The event kinematics of DIS are determined by 
$Q^2$, and one of the two 
Bjorken scaling variables $x=Q^2/2P\!\cdot\!q$
or $y = Q^2/xs,$ 
where $P$ is the four-momentum of the incoming proton
and $\sqrt s$ is the positron-proton centre-of-mass energy.
The kinematic region studied is $0.2 < y < 0.8$ and $ 0.01 < x < 0.1,$
corresponding to a $Q^2$ range $180 < Q^2 < 7220 {\rm\ GeV^2}.$

Charged particles are selected over a defined range of $p_T$ values,
where $p_T$ is the transverse momentum with respect
to the $\gamma^*P$ axis in the HCM system.
Results for the asymmetry are presented as a function of $p_c,$ the
minimum value of $p_T$ of the selected charged particles, since the 
QCD predictions depend on this variable.
The measurements are compared to
theoretical expectations.

Previous studies of single hadron production in neutral current
DIS have observed a $\cos \phi $ term that was 
attributed to non-perturbative effects~\cite{emc,e665}.
Here, for the first time, the observation of both $\cos\phi $ 
and $\cos 2\phi\ $ terms are reported. As discussed above, 
the  $\cos 2\phi\ $ term can be unambiguously attributed to perturbative
QCD processes.

\section{Experimental Setup and Event Selection}
\label{sec:expt}
The data presented here were taken at the positron-proton
 collider HERA using the ZEUS detector. The $38{\ {\rm pb^{-1}}}$
of data 
were taken in 1996 and 1997,
when HERA operated
with positrons of energy $E_e=27.5$~GeV and protons with energy
$820$~GeV.
 A detailed description of the ZEUS detector can be
found elsewhere~\cite{b:sigtot_photoprod,b:Detector}.


Throughout
this paper the standard ZEUS right-handed coordinate system is
used, in which
$X = Y = Z = 0$ is the nominal interaction point, the positive
$Z$-axis points in the direction of the proton  beam 
(referred to as the forward direction)
and the $X$-axis is horizontal, pointing towards the centre of HERA.
The polar angle, $\theta,$
is defined with respect to the positive $Z$ direction.

The key component for this analysis is the central tracking detector
(CTD)~\cite{CTD} which operates in a magnetic field of 1.43 T provided by a
thin superconducting solenoid. The CTD is a drift chamber
consisting of 72 cylindrical layers,
covering the region $15^\circ < \theta < 164^\circ$.
The transverse momentum resolution for full-length tracks is 
$\sigma(p_{T}) / p_{T} = 0.0058\, p_{T} \oplus 0.0065 $
$ \oplus 0.0014/p_{T}$ 
($p_{T}$ in GeV). 
The interaction vertex is measured using the CTD
with a typical resolution along (transverse to) the beam direction of 
0.4 (0.1) cm.

Also important for this analysis is
the uranium-scintillator sampling calorimeter (CAL)~\cite{CALRes},
which surrounds the solenoid. 
The CAL consists of 5918 cells, organised in three
sections, the forward (FCAL), barrel (BCAL) and rear (RCAL),
with longitudinal segmentation into electromagnetic and hadronic
sections.
The energy resolutions, as measured in test beams, are
$\sigma /E$ = 0.18/$\sqrt{E(\mbox{GeV})}$ and 
0.35/$\sqrt{E(\mbox{GeV})}$ for electrons and hadrons, 
respectively.

The angular coverage of the ZEUS detector allows the kinematic
variables $x$ and $y$ to be reconstructed in a variety of
ways using combinations of energies and angles
of the positron and the hadronic system.
Variables calculated only
from the measurements of the energy, $E^{\prime}_e,$
and angle, $\theta_e,$
of the scattered positron are denoted with the subscript $e$, whilst
those
calculated from the hadronic system measurements, using the 
Jacquet-Blondel method~\cite{jb}, are denoted by the subscript $JB.$
Variables calculated by these approaches
are used only in the event selection.
In the double angle method~\cite{DA}, denoted by $DA,$  the
kinematic variables are determined using $\theta_e$
and the angle $ \gamma_H $ (which is the direction of the struck
quark in the QPM), evaluated using the CAL cells corresponding to the 
hadronic final state.

The $DA$ method was used throughout this analysis
for the calculation of the boosts and the kinematic variables
because it is less
sensitive than other methods to systematic
uncertainties in the CAL energy measurement.

Corrections are applied~\cite{ZNCpaper} 
for hadronic energy loss in inactive material in front of
the calorimeter and for
the backscattering of particles into the BCAL or RCAL from hadron jets
in the FCAL direction.

The triggering and online
event selections were identical to those used for the ZEUS
measurement of the structure function $F_2$~\cite{z_shift}.

Further selection
criteria were applied both
to ensure accurate reconstruction of the kinematic
variables and to increase the purity of the sample by
eliminating background from photoproduction processes:

\begin{itemize}
\item
$E^\prime_e \ge 10~{\rm GeV}$,
to achieve a high purity sample of DIS events;
\item
$y_e\leq 0.95$,
to reduce the photoproduction background;
\item
$y_{JB}\geq 0.04$, to give sufficient accuracy for $DA$ reconstruction 
of $Q^2$ and $x$;
\item $40 \le \sum\left( E-p_Z\right)\le 60$~GeV, where the
summation is over all calorimeter cells, 
to remove photoproduction events and events with large radiative corrections;
\item $ |X|  >  16\ {\rm cm}\ {\rm or}\ |Y|  >  16\ {\rm cm} ,$ where
  $X$ and $Y$ are the impact position of the
  positron on the RCAL,
  to avoid
  the region adjacent to the rear beampipe;
\item $ -40  <  Z_{\rm vertex}  <  50\ {\rm cm},$ to reduce background
events from processes other than $ep$ collisions;
\item  $0.01 < x_{DA} < 0.1 $ and $ 0.2< y_{DA} < 0.8,$ to ensure
$\gamma_H$ lies within the CTD acceptance.
\end{itemize}

The reconstructed tracks used in the charged particle analysis
are associated with the primary event vertex
and are required to
 have $p_T^{\rm lab}>150$~MeV and $|\eta^{\rm lab}|<1.75,$
where $\eta^{\rm lab}$ is the pseudorapidity,
given by $-\ln(\tan(\theta/2)).$ 
These cuts select the region of CTD acceptance
where the detector response and systematics are best understood.

Since the gluon fragmentation function is `softer'
than that of the quarks, the hadron-quark correlation can be enhanced by
selecting `leading' charged particles. This was accomplished
by cutting on the Lorentz-invariant variable
$z_h=P\cdot p_h/P\cdot q$, where 
$p_h$ is the track or particle four-vector.
The tracks
were selected if $0.2 < z_h < 1.0;$ no selection biases
arise from
 the $\eta$ cut applied in the laboratory frame.

The number of events that pass the event selection is just over
13,800;
approximately 7,700 charged tracks satisfy the $z_h$ criteria. In
principle, more than one entry per event is permitted. 
However, as can be seen from
the above numbers, such an occurrence is rare.

\section{QCD models and event simulation}
\label{s:model}
Monte Carlo event simulation is used to correct for acceptance and
resolution effects.  The detector simulation is based on the
GEANT~3.13 program~\cite{GEANT}.

To calculate correction factors,
neutral current DIS events
were generated via DJANGO6~2.4 \cite{DJANGO},
using HERACLES~4.5.2~\cite{HERACLES}, which
incorporates first-order electroweak corrections.
The parton cascade was modelled
with the colour dipole model (CDM),
using the ARIADNE~4.10 \cite{ariadne} program, in which colour coherence
effects are implicitly included in the formalism of the parton cascade.
The Lund string fragmentation model~\cite{string} is used
for the hadronisation, as implemented in JETSET~7.4 \cite{JETSET}.
For the analysis of the data,
a Monte Carlo sample of 120k events, corresponding to an integrated
luminosity of  $67{\rm \ pb^{-1}},$ was
generated with $0.13 < y < 1.0 $ and $0.007 < x < 0.3$
and the CTEQ4D~\cite{CTEQ} parameterisation of the proton's parton
distribution functions. 
The $\rm{CTEQ4D}$ parameterisation
has been shown to describe satisfactorily the HERA measurements
of the proton structure function $F_2$ \cite{f2,h1f2} in the
kinematic regions investigated.

For systematic checks, an additional sample of 120k events
was generated using the DJANGO6 and 
LEPTO~6.5~\cite{LEPTO} Monte Carlo programs.
The parton shower option, matched to the
${\cal O}(\alpha_s)$ matrix elements (ME+PS),
was used to simulate the parton cascade.
The coherence effects in the final-state cascade are included by
angular ordering of successive parton emissions.
For the hadronisation the Lund string model was used and
for the parameterisation of the parton distribution
functions CTEQ4D was used.
The LEPTO sample was
generated with the soft colour interactions~\cite{SCI} turned off.

The ARIADNE sample was generated with the high $Q^2$ modification
developed during the 1998 HERA Monte Carlo workshop~\cite{hiQ2mod}.
A version of ARIADNE~4.10, with a corrected implementation of the $\phi$
dependence, was made
available by the author~\cite{leifblah} and it is this version that is
used in the comparison with the data.  The Pomeron-scattering option of
ARIADNE was switched off.

\section{Correction Procedure}
The Monte Carlo samples described above
were used to correct the differential $\phi$
distributions for detector effects.
To minimise the sensitivity of
the correction procedure to the underlying theoretical input
of the Monte Carlo generator, the data 
were fitted using the following approach.
Rather than
correcting the data directly, a theoretical distribution
$T(\phi_T;\vec a)$ was used 
to represent the corrected azimuthal angle, $\phi_T,$
distribution.
The parametrisation $T(\phi_T;\vec a)$ used in the
analysis is that given in equation~(\ref{eq:azi}) plus an additional
$D \sin\phi$ term, so that $\vec a$ here denotes the set of parameters
$A,B,C{\rm\ and\ } D.$
The set of parameters, $\vec a,$ was
determined by minimising the quantity

$$
\chi^2 = \sum_{\phi_M} \frac{(P_{\rm meas}(\phi_M) - P_{\rm rec}(\phi_M;\vec
a))^2}{\sigma_{M}^2},
$$

\noindent where $P_{\rm meas}(\phi_M)$ is the measured $\phi$
distribution and $P_{\rm rec}(\phi_M;\vec a)$ is  the
theoretical distribution smeared for detector effects.
The term $\sigma_M^2$ is the sum in quadrature of the uncertainties of
$P_{\rm meas}$ and $P_{\rm rec}.$
The detector smearing
takes account of: (i) migration between different $(x,y)$ bins, (ii) 
track reconstruction efficiency and the
$\phi_T$ - $\phi_M$ migration matrix and 
(iii) correction factors for pair conversion, $K^0$ decay etc.
These smearing coefficients were obtained from generated and
reconstructed Monte Carlo events.

For a bin size of $\pi/5$ radians, 
the efficiency for a generated
$\phi$ to be reconstructed in the same bin as generated
is $\sgeq 55\%.$ The purity
of the reconstructed $\phi$ bin (i.e. the percentage of the tracks
reconstructed in a bin coming from particles
 generated in the same bin) is $\sgeq 60\%.$

It should be noted that in this
procedure a corrected experimental $\phi$ distribution is not
determined.
Instead the parameterisation of the true  $T(\phi_T;\vec a)$
distribution is fitted directly to the data. 
An important feature of this method is the fact that the
$\phi_T$ - $\phi_M$ 
migration matrix depends only weakly on the generated $\phi$
distribution, and is essentially determined by the experimental
inefficiencies, which are well simulated.
Because significant off-diagonal elements are present in the matrix, this
fitting technique was used for the main analysis.

The more conventional
bin-by-bin correction
method was used as a check. Though this method is expected to be
more dependent on the underlying physics
in the generated distribution it has the advantage of producing
a corrected experimental $\phi$ distribution.
Correction factors, $F(\phi),$
were obtained from the Monte Carlo simulation by
comparing the generated distributions before the detector
and trigger simulation
with the ``observed'' distributions after these simulations followed
by the same reconstruction, selection and analysis
as the real data. The correction factor for $\phi$ is defined as
the ratio of the generated and observed
differential $\phi$ distributions:
$$
F(\phi) = \left(\frac{1}{N} \; \frac{dn}{d\phi}
\right) _{\rm gen}
 \left/ \left(\frac{1}{N} \; \frac{dn}{d\phi} \right)
_{\rm obs
} , \right.
$$
where $N$ is the number of 
Monte Carlo events and $n$ the number of charged particles
in the $(x,y)$ range. 
The correction factors are in the range 0.8-1.2. 

\section{Systematic Checks}

The major systematic errors can be divided into three types: 
uncertainties
due to event reconstruction and selection; to track selection; and to 
the modelling of the hadronic system.
No single 
systematic uncertainty was larger than the statistical error in 
the mean of either
$\cos \phi $ or  $\cos 2\phi.$
For both mean values, the largest effects, which approached the
statistical uncertainties, were associated with:
the inclusion of tracks not associated with the primary vertex; 
the use of different correction procedures for the 
energy scattered into the calorimeter from the interaction of energetic
particles in the forward direction; the use of an alternative approach to
correct for energy losses due to inactive material between the calorimeter
and the interaction
vertex; and  the $z_h$ cut being lowered by an amount commensurate with its 
reconstruction resolution.  
The mean value of $\cos 2\phi $ was also reduced by its
statistical uncertainty 
for $p_c > 1.0\ {\rm GeV}$ when LEPTO was used to correct the data.
The upper and lower uncertainties extracted from these 
systematic checks were added separately in quadrature.

Three additional cross-checks were made on the measurement that
were not included in the systematic uncertainties.
The analysis was repeated using: (i)  the bin-by-bin correction; (ii)
the measured
positron energy in place of that calculated from the DA method, to
check the sensitivity to the boost, and (iii) only 
events without bremsstrahlung from the Monte Carlo sample.
In each case the results
were consistent, within statistical errors, with those from
the original analysis.

\section{Results}

\begin{figure}[ph!]
\centering
\begin{center}
\mbox{\epsfig{file=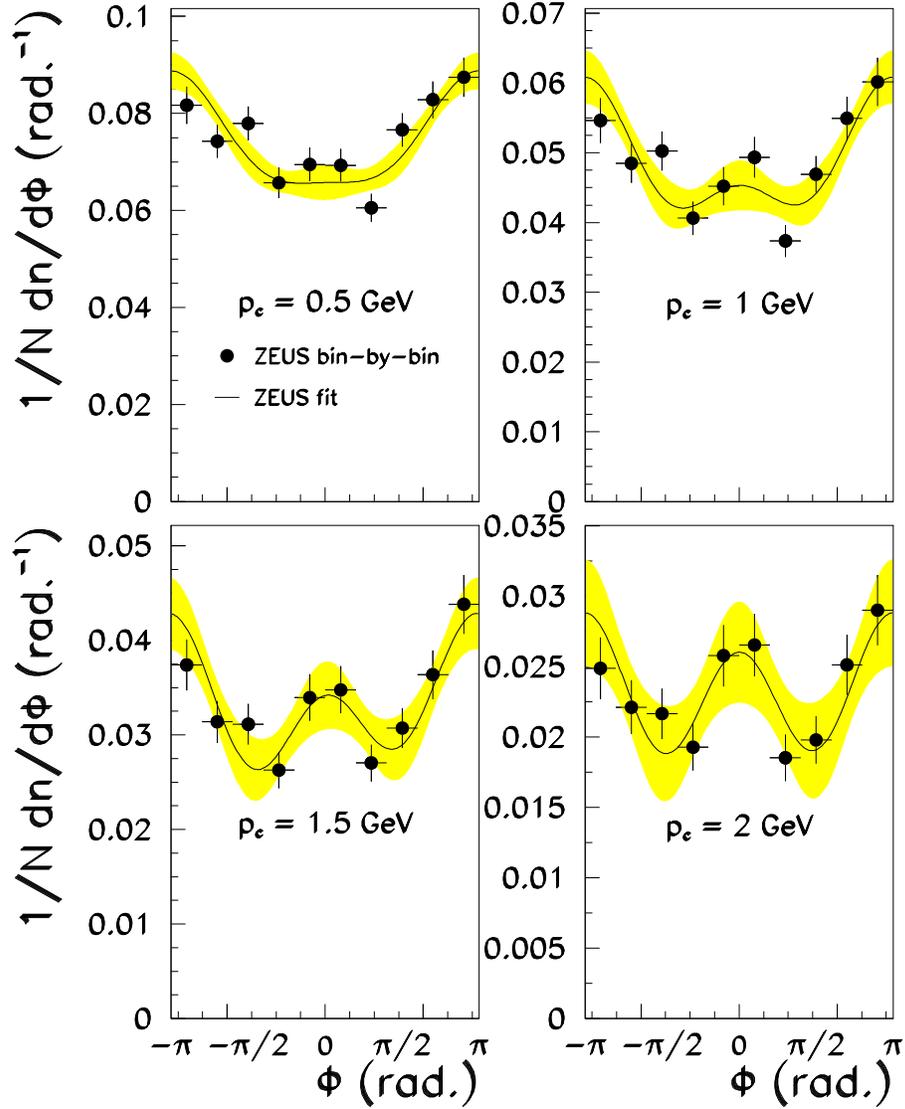,width=\textwidth}}
\end{center}
\caption{
The
differential $\phi$ distributions obtained
for four values of the $p_T$
cut, $p_c,$ in the hadronic centre-of-mass frame
in the kinematic region $ 0.01 < x < 0.1$ and $ 0.2 < y < 0.8$
for charged hadrons with $0.2 < z_h < 1.0.$
The full line
(with accompanying statistical error band) is the
result obtained from the main analysis using the 
unfolding technique. 
The data points were corrected using a bin-by-bin procedure (only
statistical errors are shown). 
}
\label{fig:phidist}
\end{figure}

Figure~\ref{fig:phidist} shows the 
differential $\phi$ distributions 
of charged hadrons 
for four values of the $p_T$
cut, $p_c,$ in the hadronic centre-of-mass.
The results obtained from the main analysis method (full line)
and the bin-by-bin
correction method (points) are seen to agree.
At low $p_c,$ a clear $\cos \phi $ term is observed. As the value of 
$p_c$ is increased a $\cos 2\phi $ term becomes evident.  

\begin{figure}[ph!]
\centering
\begin{center}
\mbox{\epsfig{file=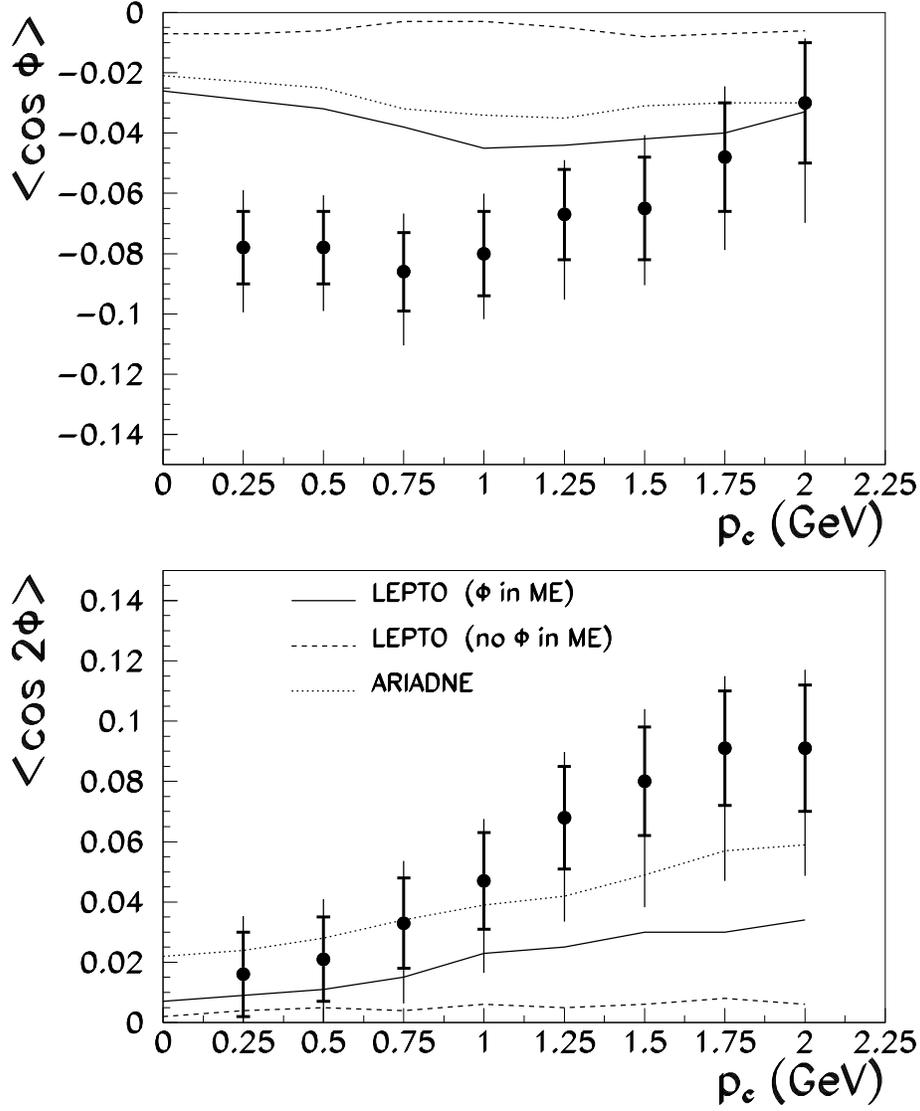,width=\textwidth}}
\end{center}
\caption{ The values of $\langle \cos \phi \rangle$
and $\langle \cos 2\phi \rangle$ are shown as a function of $p_c$ 
in the kinematic region $ 0.01 < x < 0.1$ and $ 0.2 < y < 0.8$
for charged hadrons with $0.2 < z_h < 1.0.$
The inner error bars represent the
statistical errors, the outer are statistical
and systematic errors added in quadrature.
The lines are the predictions from 
LEPTO (full line), LEPTO with flat $\phi$ dependence (dashed line)
 and ARIADNE (dotted line) Monte Carlo models.}
\label{fig:phievol}
\end{figure}

Figure~\ref{fig:phievol} shows the moments
 $\langle \cos \phi \rangle$ and $\langle \cos 2\phi \rangle$
as a function of
$p_c.$ 
They were calculated from the fit using
$
\langle \cos\phi \rangle = {B}/{2A}\ {\rm and\ }
\langle \cos2 \phi \rangle = {C}/{2A}.$
The $\sin\phi$ term
is consistent with zero independent of the value chosen for $p_c.$
The statistical errors are those obtained from fitting
the data.
The value 
of $\langle \cos \phi \rangle$ is negative and decreases in magnitude
as $p_c$ is increased. 
In contrast,
the value of $\langle \cos 2\phi \rangle$ is
positive and rises as $p_c$ is increased.
Figure~\ref{fig:phievol} also shows
predictions of
the Lund Monte Carlo programs (LEPTO and ARIADNE). 
The data have   larger absolute values of 
$\langle \cos \phi \rangle$ than predicted by LEPTO, though for higher values of $p_c$ the agreement is
reasonable. The dependence of $ \langle \cos 2\phi \rangle$
on $p_c$ is weaker in the LEPTO generator than in the data.
Imposing a flat $\phi$ dependence on the matrix elements implemented in
the LEPTO program, in order to investigate any asymmetry arising from
purely QCD coherence effects, results
in behaviour different from the data; 
both moments are almost zero 
for all values of $p_c.$
The ARIADNE Monte Carlo predictions tend to have
smaller
absolute values of $\langle \cos \phi \rangle$ than the data,
whilst the
$\langle \cos 2\phi \rangle$ term is in reasonable agreement with
the data. 

\begin{figure}[ph!]
\centering
\begin{center}
\mbox{\epsfig{file=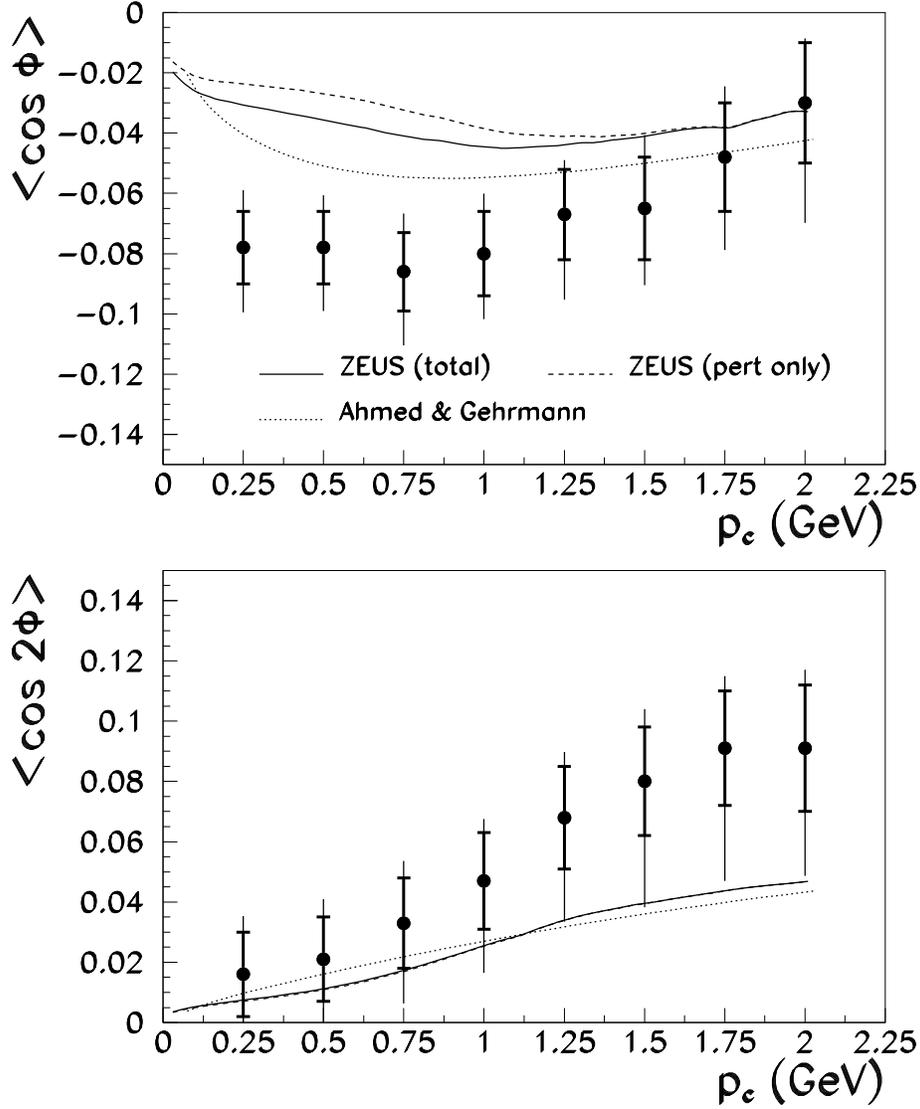,width=\textwidth}}
\end{center}
\caption{ The values of $\langle \cos \phi \rangle$
and $\langle \cos 2\phi \rangle$ are shown as a function of $p_c$ 
in the kinematic region $ 0.01 < x < 0.1$ and $ 0.2 < y < 0.8$
for charged hadrons with $0.2 < z_h < 1.0.$
The inner error bars represent the
statistical errors, the outer are statistical
and systematic errors added in quadrature.
The lines are the LO predictions from 
ZEUS with perturbative and non-perturbative 
contributions (full line),
ZEUS with the perturbative contribution only (dashed line)
and Ahmed \& Gehrmann  (dotted line -- see
text for discussion). For
the case of $\langle \cos 2\phi \rangle,$ the ZEUS total and
perturbative predictions are almost identical.}
\label{fig:mc_lo}
\end{figure}

Figure~\ref{fig:mc_lo} compares the data with two LO QCD calculations.
Both calculations were made with $Q$ as the
appropriate scale, with the Binnewies et al. LO fragmentation
function~\cite{binnewies} and with the CTEQ4
LO proton parton densities~\cite{CTEQ}.
The LO calculations result in a qualitatively similar behaviour to the LEPTO
and ARIADNE Monte Carlo generator predictions.

The analytic calculation from ZEUS (based
on the calculation of Chay et al.~\cite{chay}) 
includes an estimation of the non-perturbative
contribution, from intrinsic $k_T$ and hadronisation $p_T,$
and integrates over the whole kinematic range.
The results of Ahmed~\& Gehrmann are purely perturbative at
leading order in $\alpha_s$ and are evaluated 
at the mean values $\langle x \rangle=0.022$ and $\langle Q^2 \rangle=750\
{\rm GeV^2}$ of the data.
The different implementations
account for the observed difference in the two predictions;
using $\langle x \rangle$ and $\langle Q^2 \rangle$ in the ZEUS
perturbative calculation leads to agreement with the 
Ahmed~\& Gehrmann calculation.

There are perturbative and non-perturbative contributions to both
 $\langle \cos \phi\rangle $ and $\langle \cos 2\phi\rangle .$
For $\langle \cos\phi\rangle ,$ 
the non-perturbative contribution to the ZEUS calculation
is at most 20\% at low $p_c$ 
for a mean intrinsic $k_T$ of $0.5{\rm\ GeV}$ and a non-perturbative
fragmentation $p_T$ of  $0.5{\rm\ GeV}.$
However, this fraction depends 
on the values of $k_T$ and $p_T$ used.
For $\langle \cos 2\phi\rangle,$ 
this non-perturbative contribution is negligible, even for mean
values of
$k_T=0.9{\rm\ GeV}$ and $p_T=0.9{\rm\ GeV.}$
It can be concluded that the 
observation of a significant $\langle \cos 2\phi \rangle$ term
is clear evidence for a perturbative contribution to
the azimuthal asymmetry.

The sensitivity to higher-order corrections was examined at the
partonic level using the DISENT Monte Carlo program~\cite{DISENT}.
This study suggests that higher-order corrections are large (factors of 2-3)
at low $p_c$ for the $\langle \cos
\phi\rangle$ term, resulting in larger negative values. 
For the $\langle \cos 2\phi \rangle$ term they are much smaller.
This may explain the disagreement between the data and the LO QCD models
for $\langle \cos \phi\rangle$ at low $p_c.$

\section{Conclusions}
The azimuthal asymmetries in the deep inelastic electroproduction of single
particles have been measured at HERA in the hadronic
centre-of-mass frame in the kinematic region,  $\langle x \rangle
= 0.022$ and $\langle Q^2 \rangle =750\ {\rm GeV^2}.$
For hadrons produced at large 
transverse momenta, the measured value for  $\langle \cos\phi \rangle$
is negative and is in agreement with QCD
predictions.  
The moment  $\langle \cos 2\phi \rangle$ 
has been measured here for the first time and is
non-zero and positive.   It increases as a function of the minimum
particle transverse momentum, as expected from QCD.
Since the non-perturbative contribution to
$\langle \cos 2\phi\rangle $ is predicted to be
negligible,
this measurement provides clear
evidence for a perturbative QCD contribution to the azimuthal asymmetry.

\section*{Acknowledgements}
The strong support and encouragement of the DESY Directorate have
been invaluable, and we are much indebted to the HERA machine group
for their inventiveness and diligent efforts.  The design,
construction and installation of the ZEUS detector have been made
possible by the ingenuity and dedicated efforts of many people from
inside DESY and from the home institutes who are not listed as authors.
Their contributions are acknowledged with great appreciation.
We would like to thank Thomas Gehrmann and Mike Seymour for useful
discussions.

\newpage

\newpage

\newpage

\newpage
\end{document}